\documentclass[aps,prc,imp,twocolumn,superscriptaddress,showpacs]{revtex4}
\usepackage[pdftex,
 letterpaper=true,
 hyperindex=true,
 breaklinks=true,
 colorlinks=false,
 citecolor=blue,
 pdftitle={},
 pdfauthor={}]
{hyperref}
\usepackage{graphicx}
\usepackage{xcolor} 
\usepackage{setspace}
\usepackage[T1]{fontenc}
\usepackage{amsmath}
\begin{document}

%\title{Extracting the shape of the $\gamma$-ray strength function in short-lived nuclei}
%\title{Towards a model-independent extraction of neutron-capture reaction cross sections far from stability}
\title{A novel approach for extracting model-independent nuclear level densities far from stability}

\author{D.~M\"{u}cher}
    \email[]{dmuecher@uoguelph.ca}
	\affiliation{Department of Physics, University of Guelph, Guelph, Ontario N1G 2W1, Canada}
	\affiliation{TRIUMF, 4004 Wesbrook Mall, Vancouver, British Columbia, Canada V6T 2A3}

\author{A.~Spyrou}
    \email[]{spyrou@nscl.msu.edu}
	\affiliation{National Superconducting Cyclotron Laboratory, Michigan State University, East Lansing, Michigan 48824, USA}
	\affiliation{Department of Physics \& Astronomy, Michigan State University, East Lansing, Michigan 48824, USA}
	\affiliation{Joint Institute for Nuclear Astrophysics, Michigan State University, East Lansing, MI 48824, USA}

\author{M.~Wiedeking}
\affiliation{Department of Subatomic Physics, iThemba LABS, P.O. Box 722, Somerset West 7129, South Africa}
\affiliation{School of Physics, University of the Witwatersrand, Johannesburg 2050, South Africa}

\author{M.~Guttormsen}
\affiliation{Department of Physics, University of Oslo, NO-0316 Oslo, Norway}

\author{A.C.~Larsen}
\affiliation{Department of Physics, University of Oslo, NO-0316 Oslo, Norway}

\author{F.~Zeiser}
\affiliation{Department of Physics, University of Oslo, NO-0316 Oslo, Norway}

    \author{C.~Harris}
	\affiliation{National Superconducting Cyclotron Laboratory, Michigan State University, East Lansing, Michigan 48824, USA}
	\affiliation{Department of Physics \& Astronomy, Michigan State University, East Lansing, Michigan 48824, USA}

    \author{A.L.~Richard}
    \affiliation{National Superconducting Cyclotron Laboratory, Michigan State University, East Lansing, Michigan 48824, USA}
\affiliation{Joint Institute for Nuclear Astrophysics, Michigan State University, East Lansing, MI 48824, USA}

    \author{M.K.~Smith}
    \affiliation{National Superconducting Cyclotron Laboratory, Michigan State University, East Lansing, Michigan 48824, USA}

\author{A.~G\"{o}rgen}
\affiliation{Department of Physics, University of Oslo, NO-0316 Oslo, Norway}

\author{S.~N.~Liddick}
\affiliation{National Superconducting Cyclotron Laboratory, Michigan State University, East Lansing, Michigan 48824, USA}
\affiliation{Department of Chemistry, Michigan State University, East Lansing, Michigan 48824, USA}

\author{S.~Siem}
\affiliation{Department of Physics, University of Oslo, NO-0316 Oslo, Norway}

\author{H.~Berg}
	\affiliation{National Superconducting Cyclotron Laboratory, Michigan State University, East Lansing, Michigan 48824, USA}
	\affiliation{Department of Physics \& Astronomy, Michigan State University, East Lansing, Michigan 48824, USA}

\author{J.A.~Clark}
\affiliation{Physics Division, Argonne National Laboratory, Argonne, Illinois 60439, USA}

\author{P.A.~DeYoung}
\affiliation{Department of Physics, Hope College, Holland, Michigan 49422-9000, USA}

\author{A.C.~Dombos}
\affiliation{Department of Physics, University of Notre Dame, Notre Dame, Indiana 46556, USA}

\author{B.~Greaves}
	\affiliation{Department of Physics, University of Guelph, Guelph, Ontario N1G 2W1, Canada}

	   \author{L.~Hicks}
	\affiliation{National Superconducting Cyclotron Laboratory, Michigan State University, East Lansing, Michigan 48824, USA}
	\affiliation{Department of Physics \& Astronomy, Michigan State University, East Lansing, Michigan 48824, USA}
	
	\author{R.~Kelmar}
\affiliation{Department of Physics, University of Notre Dame, Notre Dame, Indiana 46556, USA}

	\author{S.~Lyons}
	\affiliation{Pacific Northwest National Laboratory, Richland, Washington 99352, USA}

    \author{J.~Owens-Fryar}
	\affiliation{National Superconducting Cyclotron Laboratory, Michigan State University, East Lansing, Michigan 48824, USA}
	\affiliation{Department of Physics \& Astronomy, Michigan State University, East Lansing, Michigan 48824, USA}
	
    \author{A.~Palmisano}
	\affiliation{National Superconducting Cyclotron Laboratory, Michigan State University, East Lansing, Michigan 48824, USA}
	\affiliation{Department of Physics \& Astronomy, Michigan State University, East Lansing, Michigan 48824, USA}

\author{D.~Santiago-Gonzalez}
\affiliation{Physics Division, Argonne National Laboratory, Argonne, Illinois 60439, USA}

\author{G.~Savard}
\affiliation{Physics Division, Argonne National Laboratory, Argonne, Illinois 60439, USA}

	\author{W.W.~von~Seeger}
\affiliation{Department of Physics, Hope College, Holland, Michigan 49422-9000, USA}

\date{\today}
%%%%%%%%%%%%%%%%%%%%%%%%%%%%%%%%%%%%%%%%%%%%%%%%%%%%%%%%%%%%%%%%%%%%%%%%
\begin{abstract}
The level density of quantum states in statistical mesoscopic systems is a critical input for various fields of physics, including nuclear physics, nuclear astrophysics, atomic physics and their applications. In atomic nuclei the level density is a fundamental measure of their complex structure at relatively high energies.  Here we present the first model-independent measurement of the absolute partial nuclear level density for a short-lived unstable nucleus. For this purpose we introduce the ``Shape method'' to extract the shape of the $\gamma$-ray strength function. Combining the Shape method with the existing $\beta$-Oslo technique allows the extraction of the nuclear level density without the need for theoretical input. We benchmark the Shape method using results for the stable $^{76}$Ge nucleus, finding excellent agreement to previous experimental results. We apply the Shape method to new experimental data on the short-lived $^{88}$Kr nucleus. Our method opens the door for measurements of the nuclear level density and  $\gamma$-ray strength function far away from stability, a pivotal input required to understand the role of exotic nuclei in forming the cosmos.
\end{abstract}
%%%%%%%%%%%%%%%%%%%%%%%%%%%%%%%%%%%%%%%%%%%%%%%%%%%%%%%%%%%%%%%%%%%%%%%%
 
%\pacs{26.30.-k, 26.30.Ef, 15.40.Lw }
\maketitle

%use this to add extra space for handwritten notes
%\setstretch{2}

%%%%%%%%%%%%%%%%%%%%%%%%%%%%%%%%%%%%%%%%%%%%%%%%%%%%%%%%%%%%%%%%%%%%%%%%
\noindent
\section{Introduction}

Nuclei are complex quantum many-body systems. For low-excitation energies, their structure can be described using the properties (energy, spin, parity, width) of individual levels. However, moving to higher energies these properties need to be combined into a statistical description of the nucleus, originally described in the 1930s by N.~Bohr \cite{Boh36}, as the levels get closer together and overlap. One of the most important statistical properties is the nuclear level density (NLD) as it carries information about the structure of the nucleus, such as pair breaking, shell effects, shape changes and collectivity. In addition, the NLD is a critical input in nuclear reaction calculations, in particular for neutron-capture reaction cross sections and neutron-induced fission calculations, both pivotal input for nuclear astrophysics and applications in nuclear energy and security \cite{Fon53, Raj81, Lar19, Ali06, Col10}.

Nuclear theory efforts for describing the NLD have been ongoing for more than eight decades. In 1936, H.A. Bethe first described the nucleus as a group of non-interacting fermions and showed that the NLD increases roughly exponentially as a function of excitation energy \cite{Bet36, Bet37}. Since then, numerous theoretical efforts attempted to provide a description of this important quantity, leading to modern approaches which are typically (semi)microscopic, e.g. \cite{Hun17, Gor01, Gor08}, or shell-model based, e.g. \cite{Kar16, Mus18, Orm20, Alh16, Shi16}. Theoretical calculations often have difficulty reproducing the available experimental data for stable nuclei, and their predictions diverge even more when extrapolating to regions inaccessible by experiment. Differences between models can reach an order of magnitude. This is especially important when looking at systems far from stability since these are the most sensitive regions for particular applications like for the astrophysical r process \cite{Mum16} and for nuclear energy production \cite{Ali06, Col10}.    

Experimentally, measurements of the NLD are limited to stable nuclei or their closest neighbors because the experimental approaches used to date are based on stable-beam experiments. Commonly-used techniques for extracting the NLD around stability are the Oslo method \cite{Sch00, Gut87, Gut96} and the particle-evaporation method \cite{Wal95, Voi06}. New experimental techniques were developed recently that can provide NLDs on short-lived nuclei ($\beta$-Oslo \cite{Spy14}, inverse-Oslo \cite{Ing20}). In addition, the surrogate method \cite{Rat19} was also developed recently for constraining neutron-capture reactions on unstable nuclei. However all aforementioned methods rely on inputs from theoretical models. While their results are still of high value since they are the only available methods to extract NLDs and neutron-capture reaction cross sections on short-lived nuclei, the dependence on theoretical models is a limitation. Here we present the first model-independent approach for extracting NLDs for a short-lived nucleus.

In this work we make use of the $\beta$-Oslo method \cite{Spy14} and populate the nucleus of interest via $\beta$ decay, allowing experiments with secondary beam intensities as low as 1~pps far away from nuclear stability \cite{Spy14, Lid16}. A segmented total absorption spectrometer is used to simultaneously measure the excitation energy and individual $\gamma$-ray transitions of the populated nucleus. Following the unfolding of the data with the detector response \cite{Gut96}, an iterative subtraction process allows for the extraction of the first generation $\gamma$ rays as a function of excitation energy, $E_x$ \cite{Gut87}. The extracted first generation (primary) $\gamma$-ray matrix $P(E_x, E_\gamma)$ can be factorised as \cite{Sch00}:

%%%%%%%%%%%%%%%%%%%%%%
\setlength{\mathindent}{0cm}
\begin{equation}
P(E_x,E_\gamma)\propto T(E_\gamma)\rho(E_x-E_\gamma),
\end{equation}
%%%%%%%%%%%%%%%%%%%%%%
  
 where $\rho(E_x-E_\gamma)$ is the NLD at the excitation energy after the first $\gamma$-ray is emitted, and $T(E_\gamma)$ is the transmission coefficient for $\gamma$ emission. An infinite number of solutions are possible for the above equation \cite{Sch00} and the physical solution is obtained when normalizing the $\rho(E_x-E_\gamma)$ and $T(E_\gamma)$ to known data with: 
 
 %%%%%%%%%%%%%%%%%%%%%%
 \begin{equation} \label{eqn:traf}
 \begin{split}
 \rho'(E_x-E_\gamma) & = A e^{\alpha(E_x-E_\gamma)}\rho(E_x-E_\gamma) \\
 T'(E_\gamma) & = Be^{\alpha (E_\gamma)}T(E_\gamma),
  \end{split}
 \end{equation}
%%%%%%%%%%%%%%%%%%%%%%

where $A$ and $B$ are constants and $\alpha$ is a common slope parameter. 
 Typical normalization data used in the Oslo method are: 1) low-lying discrete levels, 2) the level density at the neutron separation energy, $\rho(S_n)$, calculated from neutron-resonance spacing, $D_0$, measurements where available, and 3) the average total radiative width $\langle\Gamma_{\gamma}\rangle$ at $S_n$. Experiments performed close to the valley of stability often have reliable data for these three quantities. Whereas Coulomb dissociation measurements in inverse kinematics have been conducted \cite{Ros13, Lid16, Spy17, Lar18, Lew19} to gain a normalization of the absolute $\gamma$-ray strength in unstable nuclei, there is currently no technique to experimentally determine $\rho(S_n)$, even a few nuclei away from stability. Consequently, the absolute value of NLDs in unstable nuclei rely on theory, alone. \par

%------------------------------------------------------------------------
\begin{figure}[t!]
\includegraphics[width=1\columnwidth]{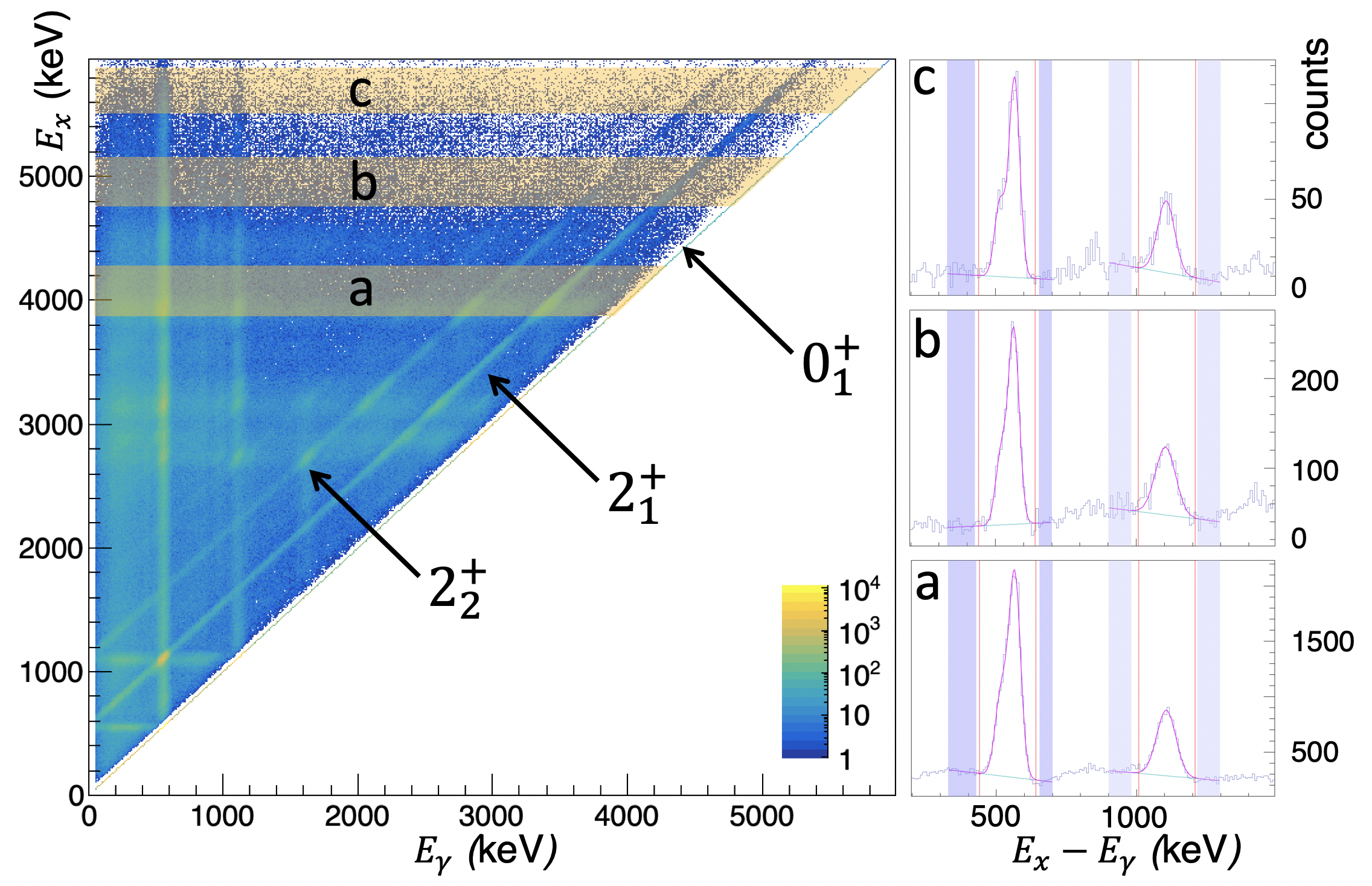}
\caption{(Color online) (left) Raw matrix of excitation energy vs $\gamma$-ray energy in SuN after population of $^{76}$Ge. (right) Projections along the diagonal, $E_x-E_\gamma$, for three different energy bins (all in keV) $3900<E_x<4400$ (a), $4700<E_x<5100$ (b), $5500<E_x<5900$ (c). The peaks at 563~keV and 1108~keV correspond to decays into the states $2^+_1$ and $2^+_2$, respectively. The purple shaded areas were used to define a linear background under each peak. }
\label{fig:matrix_ge}
\end{figure}
%------------------------------------------------------------------------

In this work, we achieve absolute normalization of the NLD by developing a new approach for extracting the shape of the $\gamma$ strength function ($\gamma$SF), a measure of the average reduced $\gamma$-ray decay probability, the so-called ``Shape method''. This method is based on the ``Ratio method'', which was proposed by Wiedeking \textit{et al.} in 2012 \cite{Wie12}, and successfully applied to $^{95}$Mo, and more recently to $^{56}$Fe \cite{Jon18} and $^{64,66}$Zn \cite{Sch20}.  The Ratio method uses particle-$\gamma$-$\gamma$ data to determine the excitation energy of the nucleus and the feeding to individual discrete levels. Through this, the dependence of the $\gamma$SF on E$_\gamma$ is determined for each excitation energy. Then the different data points that correspond to each excitation energy are combined to extract the complete shape of the $\gamma$SF. Note that the assumption is that the $\gamma$SF is independent of the excitation energy in any given nucleus for the excitation energies of interest (Brink hypothesis \cite{Bri55}).      

The Shape method relies on the observation of statistical $\gamma$ ray decays from the quasi-continuum into discrete, low-lying levels, $L_j$, with energies $E_{L_j}$. We assume that the primary $\gamma$ decays into the states, $L_j$, will be dominated by dipole transitions \cite{Kop87}. In this work we restrict the analysis to two discrete states, $L_{1,2}=2^+_{1,2}$, in the even-even daughter nucleus, i.e. two states with identical spin and parity. These transitions appear in our data as diagonals in a two-dimensional (2D) matrix with excitation energy on the y-axis and $\gamma$-ray energy on the x-axis, e.g. Fig.~\ref{fig:matrix_ge}. We use projections of the 2D matrix along the diagonals ($E_x-E_\gamma$), as shown in the same figure, for different excitation energies and extract the $\gamma$-ray intensities $N_{L_j}$ into the states of interest $L_j$. 

The ratio of the intensities $N_{L_j}$ along the diagonals, corrected for the detector response, is related to the ratio $R$ of the $\gamma$SF, $f(E_\gamma)$, for a given energy range $E_x$ \cite{Wie12}: 
\begin{equation}
R = \frac{f(E_{x,i}-E_{L_1})}{f(E_{x,i}-E_{L_2})} = \frac{N_{L_1}(E_{x,i}) (E_{x,i}-E_{L_2})^3} 
{N_{L_2}(E_{x,i})(E_{x,i}-E_{L_1})^3}.
\label{eq:slope}
\end{equation}
Equation \ref{eq:slope} can be applied for $n$ energy ranges with centroids $E_{\rm min} < E_{x,i} < Q_\beta$ and widths $\Delta E_{x,i}$ , with $E_{\rm min}$ representing the minimum energy at which the population and $\gamma$ decay of the daughter nucleus behaves statistically, and $Q_\beta$ is the Q-value of the $\beta$ decay. The $E_\gamma^3$ correction is based on the assumption that the transitions are of dipole nature.\par

The $\gamma$SF can be obtained via a ``sewing'' approach, where the pairs of data points from each excitation energy bin are  normalized to each other via linear interpolation. One specific excitation energy, $E_{x,i}$, is used as a starting point of the interpolation, yielding a pair of $\gamma$SF values at energies $E_{x,i} - E_{L_1}$ and $E_{x,i} - E_{L_2}$.  Values of the $\gamma$SF for the next higher (lower) bin are pairwise normalized such that $f(E_{x,i+1} - E_{L_2})$, and $f(E_{x,i-1} - E_{L_1}$) follow the linear trend given by the pair belonging to the bin $E_{x,i}$. The procedure is repeated for all energy bins $E_{x,i}$. Combined together, the $n$ pairs of values, $f(E_\gamma)$, reflect the shape of the $\gamma$SF of the final nucleus. Details of the sewing approach are discussed in \cite{Wie20}. \par
%------------------------------------------------------------------------
\begin{figure}[t!]
\includegraphics[width=0.95\columnwidth]{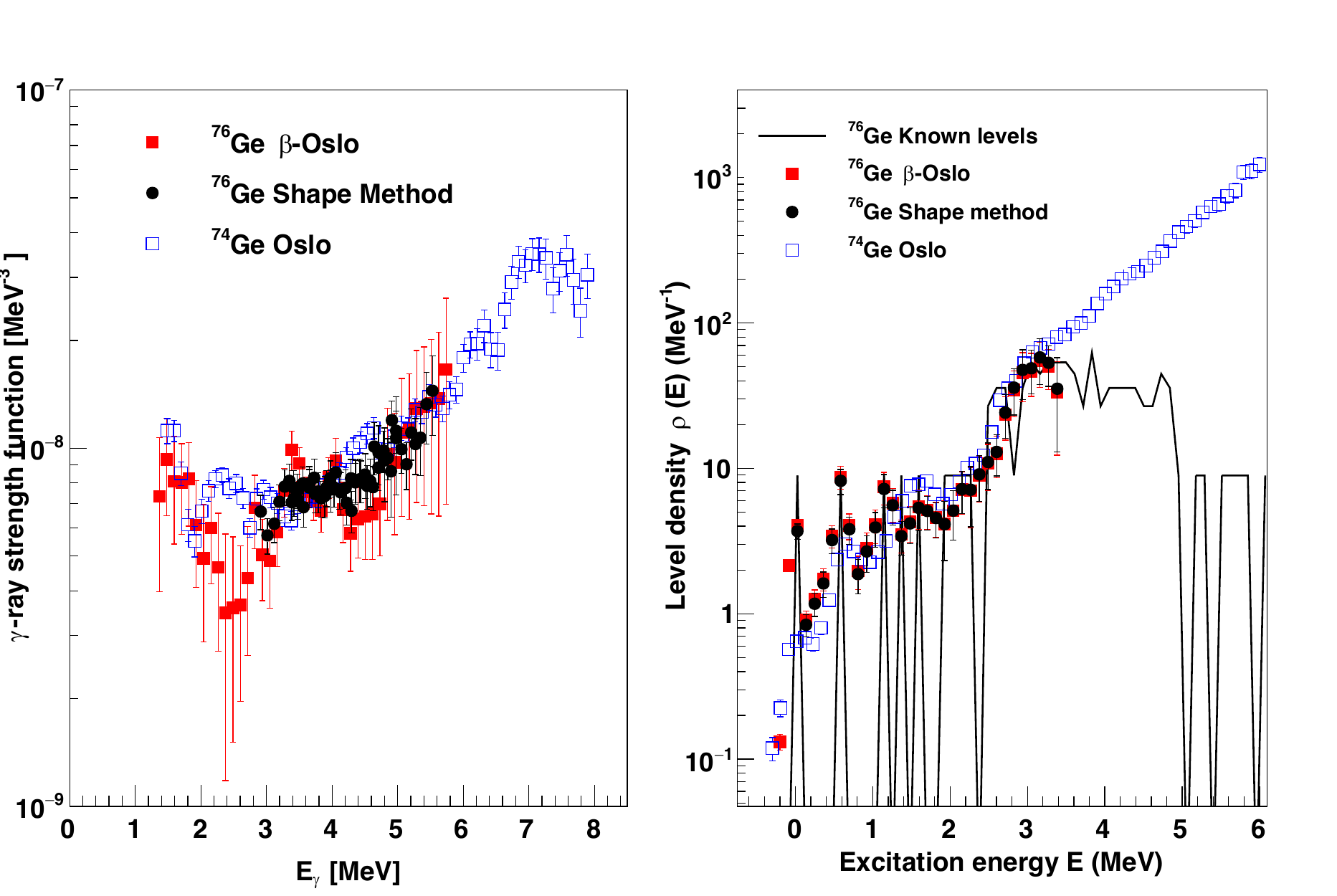}
\caption{(Color online): $\gamma$SF (left) and NLD (right) for $^{76}$Ge. Results from the original publication \cite{Spy14} for $^{76}$Ge (red squares) assume a certain level density $\rho(S_n)$ at the neutron separation energy from theory and systematics, whereas the present results (black dots) are model-independent. Results are also compared to the NLD in $^{74}$Ge extracted from the stable beam Oslo method \cite{Ren16} (blue open squares).  }
\label{fig:results-76Ge}
\end{figure}
%------------------------------------------------------------------------

%------------------------------------------------------------------------
\begin{figure*}[t!]
\begin{center}
\includegraphics[width=1.9\columnwidth]{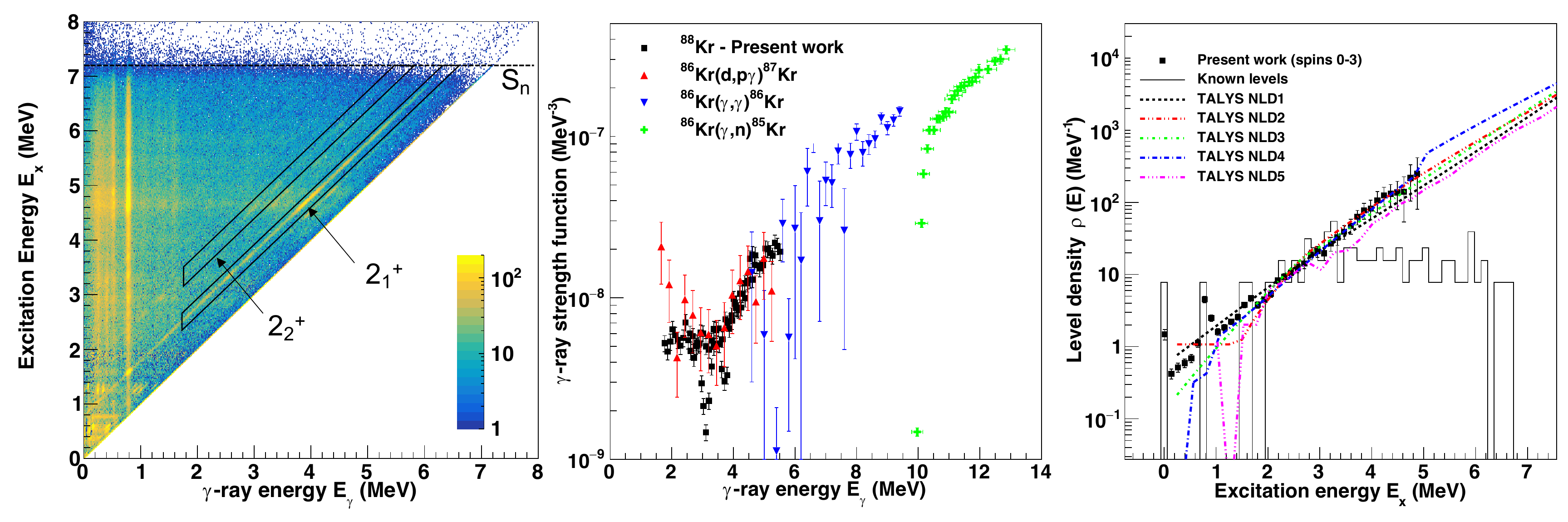}
\caption{(left) Experimental matrix of the $^{88}$Br $\beta$ decay into $^{88}$Kr. (middle) $\gamma$SF for various krypton isotopes. The black points represent the results of the present work. The red triangles are from Ref.~\cite{Ing20}, the blue inverse triangles from Ref.~\cite{Sch13} and the green crosses from \cite{Rau13}. (right) NLD of $^{88}$Kr. The data points correspond to the NLD extracted from the present work. The solid black line represents the known discrete levels, while the additional lines correspond to five NLD calculations used in the statistical model code TALYS: Constant temperature matched to the Fermi gas model (NLD1) \cite{Dil73}, Back-shifted Fermi gas model (NLD2) \cite{Dil73, Gil65}, Generalized super fluid model (NLD3) \cite{Ign79, Ign93}, Hartree Fock using Skyrme force (NLD4) \cite{Gor01} and the Hartree-Fock-Bogoliubov Skyrme force + combinatorial method (NLD5) \cite{Gor08}. }
\label{fig:Kr88-results}
\end{center}
\end{figure*}
%------------------------------------------------------------------------

In the present work the Shape Method and the $\beta$-Oslo technique are combined for the first time to extract a model-independent NLD. In parallel to this work, the Shape method has also been applied to reaction-based experiments \cite{Wie20}. To verify our approach, we have first applied the Shape method in $\beta$ decay to previously published data for the stable nucleus $^{76}$Ge, fed from the decay of $^{76}$Ga \cite{Spy14, Dom16}. The Shape method requires knowledge of the initial excitation energy of the daughter nucleus after $\beta$ decay, as well as the primary $\gamma$-ray decay into the states $L_j$ on an event-by-event basis. All data of this work make use of the Summing NaI (SuN) detector, which is a segmented total absorption $\gamma$-ray spectrometer \cite{Sim13, Dom16, Spy14}. The $^{76}$Ga secondary beams were implanted into a Si surface barrier detector at the center of SuN, providing a $\beta$-decay coincidence for the measurement of $\gamma$-rays in SuN. This is a re-analysis of an already published experiment and details can be found in the original publications \cite{Spy14, Dom16}.

Fig. \ref{fig:matrix_ge} (left) shows the raw 2D matrix, created from the SuN detector data by using the total deposited energy to get $E_x$ on the y-axis and the individual segments to get $E_\gamma$ on the x-axis (Fig.~\ref{fig:matrix_ge} (left)). The diagonals from the feeding of the first two excited states from higher lying states in $^{76}$Ge, $2^+_1$ and $2^+_2$, are clearly seen in fig.~\ref{fig:matrix_ge}, shifted by 563 and 1108~keV, respectively, from the $0^+_1$ ground state. A dedicated software ``ShapeIt'' \cite{Mue21} was developed for this analysis. 
%maybe we don't need this step? 
%The energy-calibrated matrix $(E_\gamma, E_x)$ was converted into a matrix $(E_\gamma, E_x - E_\gamma)$. In this new matrix, primary decays into states $L_j$ appear as peaks at energy $E_{L_j}$. 
Projections of the 2D matrix along the diagonal were created for $N$ bins with constant widths $\Delta E_{x_i}$ (Fig.~\ref{fig:matrix_ge} (right)). Peaks belonging to decays into $L_j = (2^+_1, 2^+_2)$ were fit using a Gaussian with a linear background. In the case of decays into the $2^+_1$ state, peaks were fit as doublets with the close-lying first-escape peak at $E_x-E_\gamma$=511~keV. For each excitation energy $E_{x,i}$, the median energy was determined and ratios of values $f(E_\gamma)$ were calculated using Eq.~\ref{eq:slope}. Applying the above described linear interpolation, the shape of the $\gamma$SF was determined. The analysis was repeated for constant energy widths 400~keV $\leq \Delta E_{x,i} \leq 800$~keV in steps of 50~keV. At each iteration, a sliding window approach was used to vary the energy of the first and all subsequent bins over the entire widths, $\Delta E_{x_i}$, in steps of 50~keV. All resulting sets of values $f(E_\gamma)$ were normalized to each other using $\chi^2$ minimization. The resulting sets of $f(E_\gamma)$ are largely independent of the choice of energy ranges, $E_{x,i}$. Details of this analysis will be discussed in \cite{Mue21}.

Figure ~\ref{fig:results-76Ge} shows the resulting $\gamma$SF of $^{76}$Ge (black dots) compared to the results of the $\beta$-Oslo method \cite{Spy14}, as well as the Oslo results for $^{74}$Ge \cite{Ren16}. It should be noted that the original publication for $^{76}$Ge using the $\beta$-Oslo method used systematics and theoretical calculations to fix the slope of the NLD and consequently the shape of the $\gamma$SF. The Shape method provides here a purely experimental approach to extracting the shape of the $\gamma$SF, in excellent agreement with the previous result. The good agreement serves as a robust test of the Shape method. The statistical uncertainties of our result are significantly reduced compared to the $\beta$-Oslo results and are comparable to uncertainties reported for the stable-beam experiment on $^{74}$Ge. Note that the Shape method can only determine the shape of the $\gamma$SF and the absolute magnitude still requires external information. Here the absolute $\gamma$SF  was adjusted to the $\beta$-Oslo result to guide the eye. 

In the following we present first results of the Shape method on an unstable nucleus, $^{88}$Kr. The new experiment was performed at the CARIBU \cite{Sav10} facility at Argonne National Laboratory. A $^{88}$Br beam was implanted into the SuNTAN tape transport system \cite{Smi20} at the center of the SuN detector decaying into the nucleus $^{88}$Kr. Isobar contaminants and daughter activity were removed from the data by using appropriate tape cycles due to their different half-lives compared to $^{88}$Br. Surrounding the implantation point, a 3~mm-thick plastic scintillator barrel was used to detect the emitted $\beta$ particles. The signals from the plastic barrel were collected by 32 wavelength-shifting optical fibers and read by two photomultiplier tubes outside of SuN. 

The 2D matrix of the populated nucleus, $^{88}$Kr, is shown in Fig.~\ref{fig:Kr88-results} (left). The $\beta$-decay Q-value of the $^{88}$Br decay is 8.97~MeV, however the data in the matrix is exhausted at roughly $E_x$ =~7~MeV which corresponds to the neutron separation energy of $^{88}$Kr ($S_n$= 7.05~MeV). The same analysis procedure that was outlined for $^{76}$Ge was applied here, once again using the diagonals corresponding to $L_j = (2^+_1, 2^+_2)$ at energies 775 and 1577~keV, respectively. The resulting $\gamma$SF is shown in Fig.~\ref{fig:Kr88-results} (middle) and is compared to other measurements of neutron-rich krypton isotopes \cite{Ing20, Sch13, Rau13}. Within the limitations of the large uncertainties in the previous measurements, the general shape of the $\gamma$SF is in excellent agreement. The observed fluctuations below about 3.5 MeV $\gamma$-ray energy are likely caused by the non-statistical behaviour of $^{88}$Kr in this low excitation energy regime. Once again, we can only compare the shape of the $\gamma$SF as the absolute scale is not constrained by our method. 

As described in the case of $^{76}$Ge, the $\gamma$SF slope $\alpha$ is used to constrain the slope of the NLD. The results are shown in Fig.~\ref{fig:Kr88-results} (right). Here we do not apply any spin corrections and instead show the model-independent experimental result for the partial NLD populated in the experiment. Assuming that the ground state of $^{88}$Br is 1$^-$ \cite{Val17} and including allowed $\beta$ transitions and dipole $\gamma$ transitions, we expect to populate spins $0-3$ of both parities. In Fig.~\ref{fig:Kr88-results} the NLD experimental results are compared to five NLD models that are commonly used in nuclear reactions applications and that are available in the open source code TALYS \cite{Kon04, Kon08}. The same spin range is used in the models as in the experiment. The agreement between experiment and theory is remarkable, showing the validity of the newly-developed method. This analysis can also be used to exclude NLD models that might not reproduce the data well enough, for example NLD5 in the case of Fig.~\ref{fig:Kr88-results}.
%------------------------------------------------------------------------
\begin{figure}[ht!]
\includegraphics[width=0.85\columnwidth]{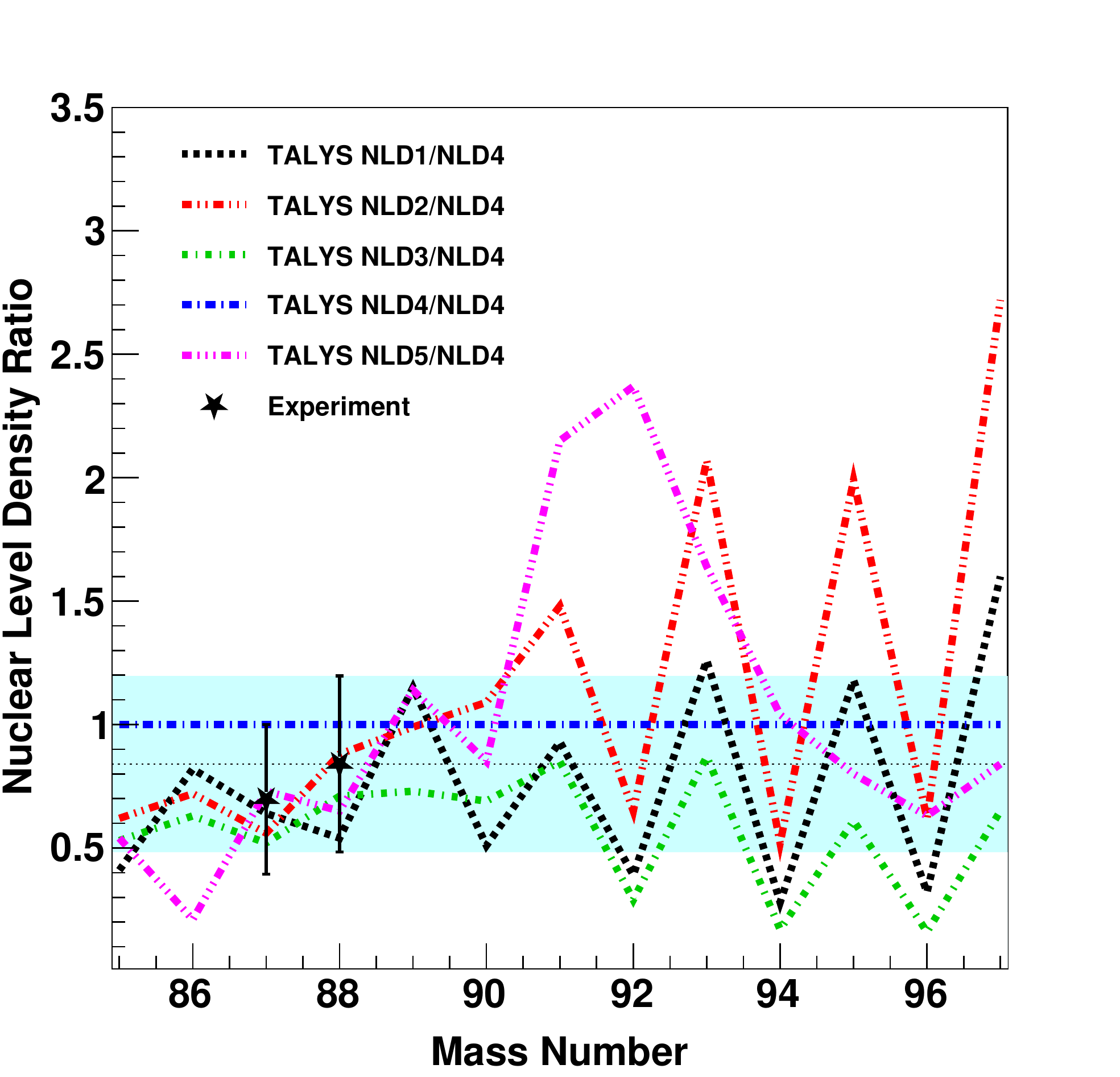}
\caption{NLD ratios of neutron-rich Kr isotopes at 4.5 MeV excitation energy using five theoretical models available in TALYS (see text for details). For comparison purposes the models are presented as a ratio against one of the available models (NLD4). The data points correspond to the NLD extracted from the present work for $^{88}$Kr and from Ref.~\cite{Ing20} for $^{87}$Kr. The experimental results are also presented as a ratio with the NLD4 model. The blue shaded area represents the $^{88}$Kr from the present work, as an indication for how our uncertainties would be able to distinguish between theoretical models for more exotic nuclei.i  }
\label{fig:NLD-all}
\end{figure}
%------------------------------------------------------------------------

$^{88}$Kr is only two neutrons away from the last stable isotope of krypton, and therefore the available NLD models are expected to be well constrained by previous experiments. However, moving into more neutron-rich isotopes, which are more critical for some of the applications mentioned earlier, the NLD models diverge significantly. This can be seen in Fig.~\ref{fig:NLD-all} which shows the ratios of the predicted NLD for the five NLD models versus model NLD4, taken at 4.5 MeV for neutron-rich isotopes of krypton in the mass range 85-97. In addition, strong differences are observed in the model predictions with respect to odd-even effects. With the technique introduced in the present work, we will be able to distinguish between models when applied to more neutron-rich isotopes. 

In summary, we introduced in this work and in \cite{Wie20} the Shape method, which is a new technique that can provide the shape of the $\gamma$SF in a model-independent way. In combination with the $\beta$-Oslo method, we were able to extract a model-independent NLD for the partial spin range populated in $^{76}$Ge and unstable $^{88}$Kr. Thanks to the sensitivity of the $\beta$-Oslo technique in combination with the Shape method our error bars are comparable to those achieved in stable beam experiments. This opens up a new avenue to study the partial NLDs of thousands of unstable nuclei far away from stability, with major impacts on our understanding of nuclear structure, nuclear astrophysics and nuclear applications. Specifically, our results will allow for significantly reduced model-dependence and uncertainties in constraining neutron-capture rates of neutron-rich nuclei via the $\beta$-Oslo and inverse-Oslo technique. This will allow the acquisition of highly-demanded information on neutron-capture rates in the r process \cite{Mum16}. We also plan to investigate the possible application of the Shape method to atomic physics of heavy elements where level densities were found to behave statistically and similarly to NLDs towards the ionization energy \cite{Fla94}, with impact on e.g. the development of the nuclear $^{229}$Th clock \cite{Dzu10}.

%Acknowledgements
\begin{acknowledgments}

The authors acknowledge support of the operations staff at the National Superconducting National Laboratory and at the ATLAS facility at Argonne National Laboratory. This research was partially supported by the Natural Sciences and Engineering Research Council (NSERC) of Canada. The work was supported by the National Science Foundation under grants
PHY 1913554,  %Window on the Universe
PHY 1350234, %CAREER
PHY 1430152, %JINA 
PHY 1565546, %NSCL
PHY 1613188. %Hope
 This material is based upon work supported by the Department of Energy/National Nuclear Security Administration through the Nuclear Science and Security Consortium under Award No. DE-NA0003180. A.~C.~L. gratefully acknowledges funding by the European Research Council through ERC-STG-2014 under grant agreement 637686, and support from the ``ChETEC'' COST Action (CA16117), supported by COST (European Cooperation in Science and Technology). This work is based on the research supported in part by the National Research Foundation of South Africa (Grant Number 118846). This material is based upon work supported by the U.S. Department of Energy, Office of Science, Office of Nuclear Physics, under contract number DE-AC02-06CH11357. S.L. was supported by the Laboratory Directed Research and Development Program at Pacific Northwest National Laboratory operated by Battelle for the U.S. Department of Energy. This work is partly supported by the Research Council of Norway (Grant Number: 263030).
\end{acknowledgments}

%

%\bibliographystyle{apsrev4-1}
%\bibliography{references}
\end{document}